\documentstyle[11pt,epsf,paspconf]{article}

\begin{document}

\title{What is Beta?}

\author{Andreas A. Berlind, Vijay K. Narayanan \& David H. Weinberg}
\affil{Astronomy Department, The Ohio State University, Columbus, OH 43210}



\begin{abstract}
Measurements of the cosmological density parameter ($\Omega$) using techniques
that exploit the gravity-induced motions of galaxies yield, in linear 
perturbation theory, the degenerate parameter combination $\beta = 
\Omega^{0.6}/b$, where the linear bias parameter $b$ is the ratio 
of the fluctuation amplitudes of the galaxy and mass distributions. However, 
the relation between the mass and the galaxy density fields depends on the 
complex physics of galaxy formation, and it can in general be non-linear, 
non-local, and stochastic. There is a growing consensus that the 
one-parameter model for bias is oversimplified.  This leaves us with the 
following question: What is the quantity that is actually being measured by 
the different techniques?  In order to address this question, we present 
estimates of $\beta$ from galaxy distributions constructed by applying a 
variety of biased galaxy formation models to cosmological N-body simulations.
We compare the values of $\beta$ estimated using two different techniques:
the anisotropy of the redshift-space power spectrum and an idealized version
of the POTENT method.  In most cases, we find that the bias factor 
$b = \Omega^{0.6}/\beta$ derived from redshift-space anisotropy or POTENT 
is similar to the large-scale value of $b_{\sigma}(R)$ defined by the ratio 
of rms fluctuation amplitudes of the galaxy and mass distributions.
However, non-linearity of bias and residual effects of non-linear 
gravitational evolution both influence $\beta$ estimates at the 10-20\% level.
\end{abstract}

\section{Analysis}
We estimate $\beta$ from a variety of simulated galaxy distributions, 
created by applying simple local and deterministic biasing algorithms 
to the mass distributions obtained from cosmological N-body simulations.  

The {\bf cosmological models} that we use are:
(1) $\Omega=1.0$, CDM model with a tilted power spectrum designed to 
simultaneously satisfy both COBE and cluster normalization constraints.
(2) $\Omega=0.4$, CDM, cluster normalized model.
(3) $\Omega=0.2$, CDM, cluster normalized model.
Cluster normalization requires that $\sigma_{8,m}\Omega^{0.6}=0.55$
(White, Efstathiou \& Frenk 1993).

The {\bf biasing algorithms} that we use are:
(1) Semi-analytic: An empirical 
bias prescription derived by Narayanan et al. (1999, in prep.) which 
characterizes the relation between the galaxy and mass density fields
in the semi-analytic galaxy formation models of Benson et al. (1999). 
(2) Sqrt-Exp. (Square-root Exponential): A biasing prescription in which
$(1+\delta_g) \propto \sqrt{(1+\delta_m)}e^{\alpha(1+\delta_m)}$.
(3) Power-law: A biasing prescription in which $(1+\delta_g) \propto 
(1+\delta_m)^{\alpha}$.  Here, $(1+\delta_g)$ and $(1+\delta_m)$ are the
galaxy and mass over-densities, respectively.

All the bias models are designed to yield galaxy distributions with an 
rms fluctuation, in top-hat spheres of radius $12 h^{-1}$Mpc, of 
$\sigma_{12}\approx0.7$.  The values of $\Omega^{0.6}/b_{\sigma}(12)$ for 
the galaxy distributions are 0.6, 0.58, and 0.56 for $\Omega=1.0$, 0.4, 
and 0.2, respectively.  All the results have been averaged over four 
independent realizations of the cosmological models.  The simulation
volumes are periodic cubes, with a box size of $400 h^{-1}$Mpc.

The first method that we use to estimate $\beta$ involves the anisotropy
of the redshift-space power spectrum.
Kaiser (1987) and Cole, Fisher \& Weinberg (1994) showed that in linear 
theory, and under the plane-parallel approximation, the multipole moments of 
the redshift-space power spectrum may be used to estimate $\beta$.  Here we 
employ two different $\beta$ estimators: $P^S(k)/P^R(k)$,
the ratio of the angle-averaged redshift-space power spectrum (monopole) to 
the real-space power spectrum, and $P_2(k)/P_0(k)$, 
the ratio of the quadrupole and monopole moments of the redshift-space power 
spectrum.  These ratios are, in principle, measurable, and they are related 
to $\beta$ as follows:
$\frac{P^S(k)}{P^R(k)} = 1+\frac{2}{3}\beta+\frac{1}{5}\beta^2$,
$\frac{P_2(k)}{P_0(k)} = (\frac{4}{3}\beta+\frac{4}{7}\beta^2)/
(1+\frac{2}{3}\beta+\frac{1}{5}\beta^2)$.
These ratios yield an estimate of $\beta$ at each wavenumber $k$, and
the scale dependence of this estimate is caused mainly by non-linearity in the
peculiar velocity field.  It is possible to obtain a global estimate of 
$\beta$ by modeling this non-linearity.  We use two such models in our
analysis: (1) Exponential velocity distribution model (Cole, Fisher \& 
Weinberg 1995), where we assume that the galaxies have uncorrelated 
small scale peculiar velocities which are drawn from an exponential 
distribution.  We use this model in our analysis of 
$P^S(k)/P^R(k)$ (the results of which are shown in Fig. 1).
(2) Empirical model (Hatton \& Cole 1999), where the scale dependence 
of $P_2(k)/P_0(k)$ is found empirically by examining a large number of
N-body simulations spanning a broad range of parameter space.

The second method that we use to estimate $\beta$ involves a direct 
comparison between the mass and galaxy density fields.
The POTENT method (Bertschinger \& Dekel 1989) reconstructs
the full three dimensional peculiar velocity field ${\bf v}$ from the observed
radial peculiar velocity field, allowing a measurement of $\beta$ directly 
from the linear theory relation $\nabla \cdot {\bf v} = - \beta H \delta_g$.
We estimate $\beta$ using an idealized POTENT method, in which we have
perfectly reconstructed the three dimensional peculiar velocity field.
We obtain the value of $\beta$ by fitting a line of the form $y=ax+c$ to the
$-(\nabla \cdot {\bf v})$ vs. $\delta_g$ relation.  We perform this fit
in the region: $-0.5 < \delta_g < 0.5$.  By repeating this procedure for
different smoothing of the velocity and density fields, we
can measure $\beta$ as a function of scale (see Fig. 1).

\section{Conclusions}

Figure 1 compares $\beta$ estimates from the two methods described
previously to the function $\beta_{\sigma}(R) = \Omega^{0.6}/b_{\sigma}$. 
Here, $b_{\sigma}(R)$ is the bias function defined as 
$b_{\sigma}(R) = \sigma_g(R)/\sigma_m(R)$, where $\sigma_g(R)$ and 
$\sigma_m(R)$ are the rms fluctuations of the galaxy and mass density
fields, smoothed with a Gaussian filter of radius $R$, 
of the galaxy and mass distributions, respectively.
In our full analysis, we also compare our $\beta$ estimates to the 
function $\beta_P(k) = \Omega^{0.6}/b_P(k)$.  Here $b_P(k)$ is the
bias function defined as $b_P(k) = \sqrt{P_g(k)/P_m(k)}$,
where $P_g(k)$ and $P_m(k)$ are the power spectra of the galaxy and 
mass distributions, respectively.  In the case of a simple, linear bias 
model, $\delta_g=b\delta_m$, all these definitions of $b$ are equivalent, 
i.e., $b_{\sigma} = b_P = b$.

In most cases, we find that the bias factor $b = \Omega^{0.6}/\beta$
derived from redshift-space anisotropy or POTENT is similar to the large-scale
value of $b_P(k)$ or $b_{\sigma}(R)$ defined by rms fluctuation amplitudes.
Discrepancies at the 10-20\% level may reflect non-linear dynamics as much
as non-linearity of bias, since they also occur for unbiased models.
In greater detail:
(1) Measurements of $\beta$ derived by fitting the Cole et al. (1995)
exponential velocity distribution model to $P^S(k)/P^R(k)$
underestimate the large-scale values of $\beta_P$ by about 10-20\% for all 
cosmological models and biasing prescriptions.
(2) Measurements of $\beta$ derived by fitting the Hatton \& Cole (1999)
empirical model to $P_2(k)/P_0(k)$ accurately reproduce the large-scale 
values of $\beta_P$ for all cosmological models and biasing prescriptions.
(3) Measurements of $\beta(R)$ derived from a POTENT-like comparison
of $-(\nabla \cdot {\bf v})$, the negative divergence of the peculiar
velocity field, to $\delta_g$, the galaxy density field, underestimate
the large-scale values of $\beta_{\sigma}$ by about 10-20\% for all
cosmological models and biasing prescriptions, with the exception of the
Sqrt-Exp. biasing prescription.  The Sqrt-Exp. model gives erratic
results because the non-linearity of the bias model makes the estimated
$\beta$ sensitive to the range of $\delta_g$ used in the linear fit.
(4) Estimates of $\beta$ from the anisotropy of the redshift-space 
power spectrum approximately agree with those from the POTENT method on 
large scales, except in the case of the Sqrt-Exp. model.

\begin{figure}
\centerline{
\epsfxsize=5.25truein
\epsfbox[18 242 591 705]{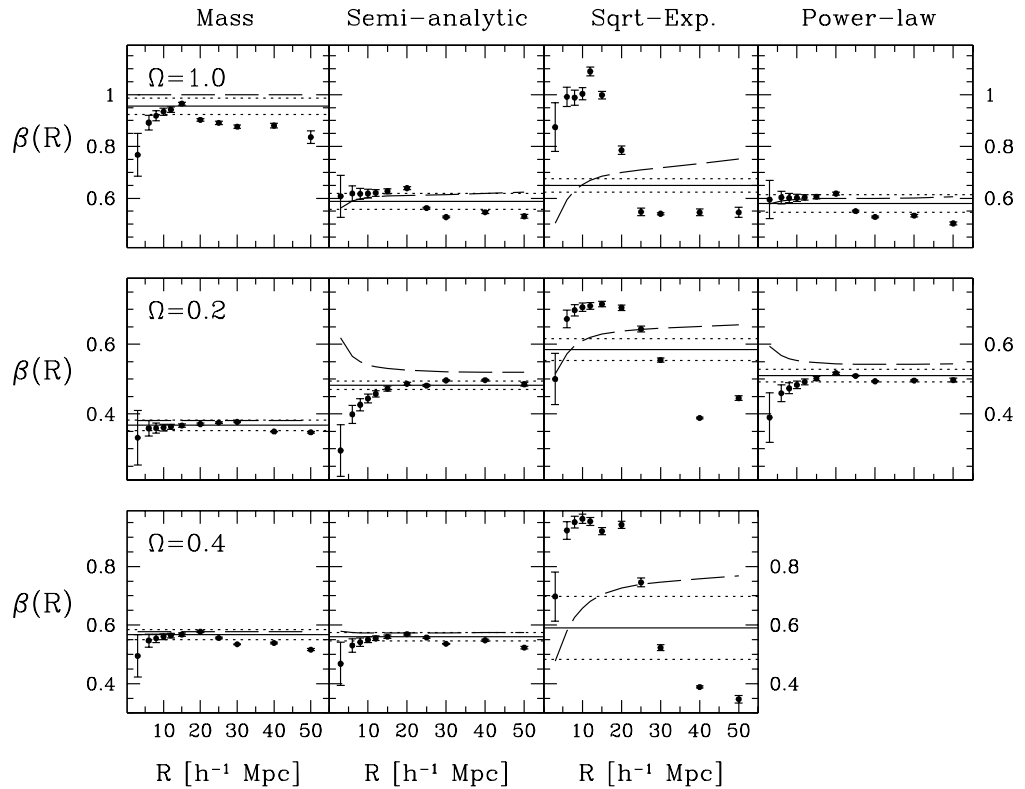}
}
\caption{
Comparison of $\beta$ estimates from different methods, as a function of 
scale.  Each panel shows $\beta(R)$ for a particular cosmology and biasing 
prescription.  The dashed line represents the function $\beta_{\sigma}(R) = 
\Omega^{0.6}/b_{\sigma}(R)$.  Here, $b_{\sigma}(R)$ is the bias function
defined as $b_{\sigma}(R) = \sigma_{g}(R)/\sigma_{m}(R)$, where 
$\sigma_{g}(R)$ and $\sigma_{m}(R)$ are the rms fluctuations of the galaxy
and mass density fields, smoothed with a Gaussian filter of radius $R$.
The solid and dotted lines show the estimates of $\beta$, along with their 
$1\sigma$ uncertainties, derived from fitting the exponential velocity
distribution model (Cole et al. 1995) to $P^S(k)/P^R(k)$.
The points represent POTENT-like $\beta$ estimates, derived by fitting 
a line to the measured relation between $-(\nabla \cdot {\bf v})$ and $\delta_g$, 
when the velocity and density fields are smoothed with Gaussian filters of 
different radii, $R$.  Each point represents the weighted average over 
four independent simulations, and its $1\sigma$ error is computed as the
error in the weighted mean.  With the exception of the galaxy distributions 
created using the Sqrt-Exp. biasing prescription, the large-scale values
of $\beta$ derived from the $-(\nabla \cdot {\bf v}) - \delta_g$ relation 
are similar to the large-scale values of $\beta_{\sigma}$, underestimating
them by about 10-20\%.  Furthermore, these estimates are in approximate 
agreement with $\beta$ estimates derived from the anisotropy of the
redshift-space power spectrum.
} \label{fig-1}
\end{figure}

\end{document}